\documentclass[aps,twocolumn,nofootinbib,showpacs, prl]{revtex4-2}

\usepackage{hyperref}
\usepackage{amsmath,amsfonts,amssymb}
\hypersetup{dvips,dvipdfm,colorlinks=true,urlcolor=magenta,filecolor=magenta,linktoc=page,citecolor=red,linkcolor=blue,bookmarks=true}
\usepackage{graphicx,epstopdf}
\usepackage{multirow}
\usepackage{xcolor}
\begin{document}
	\title{Classical and quantum analysis of gravitational singularity from Raychaudhuri equation}
	\author{Madhukrishna Chakraborty ~and~
		Subenoy Chakraborty*}
	\affiliation{Department of Mathematics, Jadavpur University, Kolkata - 700032, India}
	\begin{abstract}
		The present work deals with the Raychaudhuri equation (RE) to examine the space-time singularity both from classical and quantum point of view. The RE has been looked upon in terms of a classical linear Harmonic oscillator and Focusing theorem has been studied. Also, the Raychaudhuri scalar has been shown to affect the convergence and singularity formation in Black-holes and cosmology.  Finally, resolution of initial big-bang singularity has been attempted quantum mechanically in two ways namely by canonical quantization and by formulation of  Bohmian trajectories.
	\end{abstract}
	\maketitle
	\section{I.Introduction} The appearance of singularity in Einstein's general theory of relativity is a well-known fact. Although after the detection of gravitational waves, Einstein's general theory of relativity is the most widely accepted theory of gravity to describe physical reality yet it did not take much time for the cosmologists and relativists to realize that even the two simplest space-time solutions of the Einstein's field equations namely the Schwarzschild metric and the Friedmann–Lemaître–Robertson-Walker (FLRW) metric, bear gravitational singularity.  At the singularity, there is no structure of space-time and physical laws break down. Besides being just a matter of mathematical curiosity, such singularities depict the beginning of the Universe at the Big Bang and the terminal point of gravitational collapse in black holes as well. So a natural question arises-- ``Is this singularity inevitable in general relativity?"
	
	Raychaudhuri in the early 1950's tried to address this issue by formulating an evolution equation for expansion scalar---the Raychaudhuri equation (RE) \cite{raychaudhuri},\cite{Raychaudhuri:1953yv},\cite{Dadhich:2005qr},\cite{Dadhich:2007pi},\cite{Ehlers:2006aa},\cite{Kar:2008zz},\cite{Kar:2006ms},\cite{Chakraborty:2023rgb}. In the recent times, the RE has been extensively discussed and analyzed in a variety of frameworks of general relativity, string theory, the theory of relativistic membranes, and in a vast range of physical contexts like in gravitational lensing, crack formation in spherical astrophysical objects etc. In his seminal paper published in 1955, Raychaudhuri first pointed out the connection of his equation to the existence of singularity. Subsequently, after a decade Hawking and Penrose formulated the seminal singularity theorems in General Relativity (GR) using this RE as the main ingredient through the notion of geodesic focusing \cite{Penrose:1964wq},\cite{Hawking:1970zqf},\cite{Hawking:1973uf}.  In this context Landau and Lifschitz expressed that a singularity always implies focusing of geodesic but focusing alone can not imply a singularity. The essence of the original 1955 paper can be found in a review work by Kar and SenGupta (for ref. see \cite{Kar:2006ms}). The major application of RE lies in studying gravitational focusing properties in cosmology. When a star is heavier than a few solar masses, it undergoes an endless gravitational collapse without achieving any equilibrium state. The final outcome of a massive star must necessarily be a black hole which covers the resulting space-time singularity and any external observer at infinite can not receive a causal message from the singularity. Resolution of these singularities have remained a puzzle over the decades.

	Given a matter with an equation of state, the scale factor $a(t)$ and its time derivatives (giving rise to various cosmographic parameters) encode all dynamics which can be obtained from the RE. The classical or modified RE helps to change the sign of $\ddot{a}$ by choosing appropriate matter. For example in classical cosmology if matter obeys (does not obey) Strong Energy Condition (SEC) then RE yields $\ddot{a}<0~(>0)$. However in general for extended theories of gravity the repulsive/ more attractive nature of gravity deduced from the RE is not restricted to the choice of matter. Due to an interplay between matter and gravitational sectors, RE in modified gravity theories may take a form such that $\ddot{a}$ can become positive even for matter which does not violate the SEC and vice-versa. In this context, it is worthy to discuss about the fact that the RE describes the evolution of a congruence in a space-time and is purely geometrical. One does not need to assume any gravity theory and this gives the freedom to use any theory to fix the geometry and then study the evolution of congruence.

	It is generally speculated that a quantum description of gravity may resolve these issues. Though there is no universally accepted theory of quantum gravity, there are at present two major approaches for formulating a quantum theory of gravity-- canonical quantization \cite{DeWitt:1967yk} and path integral formulation \cite{Hawking:1978jz}. Implication of the RE in quantum background may be helpful in determining the existence or resolution of singularity at the quantum level. In this context, Das \cite{Das:2013oda} formulated a quantum version of the RE using Bohmian trajectories and argued that Convergence Condition may be avoided in the presence of quantum potential (for ref see \cite{Das:2013oda}). Also RE has been explored in the context of avoidance of black hole singularity in loop quantum cosmology \cite{Ashtekar:2008ay}. Moreover RE has also been investigated in the space-time described by the qmetric and it has been proved that the existence of the zero point length (included in a qmetric) can possibly avoid the focusing of geodesic \cite{Chakraborty:2019vki}. Motivated by these works in literature, in the present work, we show two different ways to avoid the classical singularity: (i) by quantizing the classical geodesic flow followed by Wheeler-DeWitt quantization scheme (WD formalism) and (ii) by replacing the classical geodesics with quantum Bohmian trajectories (Bohmian Formalism).

	The plan of the letter is as follows: Section II deals with the basics of the Raychaudhuri equation and Convergence Condition in terms of the frequency of a classical linear harmonic oscillator. Section III discusses the implication of the Raychaudhuri scalar in forming Black-hole and cosmological singularity. Section IV shows the quantum resolution of classical singularity via Lagrangian and Hamiltonian formulation of RE giving rise to two formalisms namely, WD formalism and Bohmian formalism.  The letter ends with conclusion and discussion of the obtained results in Section V.
	\section{II. Raychaudhuri Equation and Focusing Theorem: A linear Harmonic Oscillator approach}
	The kinematics of flows in a geometrical space are characterized by the Raychaudhuri equation. In general, a vector field generates flows and the integral curves (which may be geodesics or non geodesic congruences) of the vector field lead the flow. In Lorentzian space these congruences may be either time-like or null in nature. The evolution of the kinematics (which characterize the flow) are in general known as Codazzi- Raychaudhuri equation. However, historically only the evolution of the expansion scalar is termed as the RE. 
	Let $u^{a}$ be the velocity vector field along the integral curves ($u_{a}u^{a}=-1$ for time-like curves) then its gradient is a second rank tensor which can be decomposed into the following form
	\begin{equation}
		B_{ab}=\nabla_{b}u_{a}=\sigma_{ab}+\omega_{ab}+\dfrac{1}{n-1}q_{ab}\Theta.
	\end{equation} Here \begin{equation}	\small \sigma_{ab}=\sigma_{(ab)}=\frac{1}{2}(\nabla_{a}u_{b}+\nabla_{b}u_{a})-\frac{1}{n-1}~q_{ab}~\Theta+\frac{1}{2}~(A_{b}u_{a}+A_{a}u_{b})
	\end{equation} is the shear tensor (a symmetric traceless component of $B_{ab}$), 
	\begin{equation}\omega_{ab}=\frac{1}{2}(\nabla_{b}u_{a}-\nabla_{a}u_{b})-\frac{1}{2}(A_{b}u_{a}-A_{a}u_{b})
	\end{equation} is the vorticity tensor (anti-symmetric component of $B_{ab}$), \begin{equation}\Theta=\nabla_{a}u^{a}\end{equation} 
	is known as the expansion scalar ( the trace of $B_{ab}$ which describes how congruence focus or defocus), \begin{equation}A_{a}=(\nabla_{b}u_{a})u^{b}
	\end{equation} is the acceleration vector, \begin{equation}q_{ab}=g_{ab}+u_{a}u_{b}
	\end{equation} is the projection tensor (induced metric) and $n$ is the dimension of the space-time. Here $\sigma_{ab}$ and $\omega_{ab}$ are purely spatial as $u^{a} \sigma_{ab}=0=u^{a}\omega_{ab}.$ Then the RE i.e the evolution of the expansion scalar along the flow described by a time-like congruence is a first order differential equation as 
	\begin{equation}\label{eq2}
		\dfrac{\mathrm{d}\Theta}{\mathrm{d}\tau}=-\frac{1}{n-1}\Theta^{2}-2\sigma^{2}+2\omega^{2}+\nabla_{b}A^{b}-\tilde{R},
	\end{equation} where $2\sigma^{2}=\sigma_{ab}\sigma^{ab}$, $2\omega^{2}=\omega_{ab}\omega^{ab}$, $\tilde{R}=R_{ab}u^{a}u^{b}$ is termed as  Raychaudhuri scalar and the parameter $\tau$ identifies the proper time. Geometrically $\tilde{R}$ can be interpreted as mean curvature in the direction of $u$ \cite{Albareti:2012se}. From mathematical point of view the RE can be termed as Riccati equation and it becomes a linear second order equation as 
	\begin{equation}
		\dfrac{d^{2}Y}{d\tau^{2}}+\omega_{0}^{2}~Y=0,\label{eq8}
	\end{equation} where $\Theta=(n-1)\dfrac{d}{d\tau}\ln Y$, $\omega_{0}^{2}=\dfrac{1}{n-1}(\tilde{R}+2\sigma^{2}-2\omega^{2}-\nabla_{b}A^{b}).$ Thus the RE can be identified as a linear harmonic oscillator equation with time varying frequency $\omega_{0}$. As 	$\Theta$ may be defined as the derivative of the geometric entropy ($S$) or an average (or effective)
	geodesic deviation so one may identify $S = lnY$. The expansion $\Theta$ is nothing but the rate of change of volume of the transverse subspace of the congruence/bundle of geodesics. Therefore, the expansion approaching negative infinity (i.e. $\Theta\rightarrow-\infty$) implies a convergence of
	the bundle, whereas a value of positive infinity (i.e. $\Theta\rightarrow+\infty$) would imply a complete divergence. Thus the
	Convergence Condition (CC) can be stated as follows:
	\begin{enumerate}
		\item 	(i) Initially $Y$ is positive but decreases with proper time i.e $\dfrac{dY}{d\tau}<0$.

		\item 	(ii)  Subsequently $Y=0$ at a finite proper time to have negative infinite expansion.
	\end{enumerate}
	From the above interrelation: $\Theta=\dfrac{(n-1)}{Y}\dfrac{dY}{d\tau}$, it is clear that there should be an initially negative expansion (i.e. $\Theta(\tau=0)<0$) and subsequently $\Theta\rightarrow-\infty$ as $Y\rightarrow0$ at a finite proper time. Therefore the CC essentially coincides with the condition for the existence of zeroes of $Y$ in finite proper time. However the Sturm comparison theorem (in differential equation) shows that the existence of zeros in $Y$ at finite value of the proper time $\tau$ requires
	\begin{equation}
		(\tilde{R}+2\sigma^{2}-2\omega^{2}-\nabla_{c}A^{c})\geq0\label{eq34*}.
	\end{equation} Therefore (\ref{eq34*}) is the CC for a congruence of time-like curves (may be geodesic or non geodesic). Further, the above inequality shows that, the Raychaudhuri scalar $\tilde{R}$ and the shear/anisotropy scalar $2\sigma^{2}$ are in favor of convergence of the congruence of time-like curves while rotation and acceleration terms oppose the convergence. The CC reduces to $\tilde{R}\geq0$ if we consider the congruence of time-like
	curves to be geodesic and orthogonal to the space-like hyper-surface. This leads to \textbf{Geodesic
		Focusing} and hence the \textbf{Focusing Theorem}. In other words, rotation and acceleration terms act against the focusing but shear and
	Raychaudhuri scalar are in favor of it. Thus from physical point of view, if the RE corresponds to a realistic linear harmonic oscillator then it is inevitable to have a singularity. We name the scalar: $R_{c}=\tilde{R}+2\sigma^{2}$, as the Convergence scalar.
	Since in other modified theories of gravity the field equations are different, there may arise certain conditions for the possible avoidance of singularity even with the assumption of Strong Energy Condition (SEC) \cite{Chakraborty:2023ork}. Therefore in the context of avoidance of classical singularity, the role of RE is also equally important in modified gravity theories \cite{Choudhury:2021zij}.

	Further in the homogeneous and isotropic FLRW background ($\sigma^{2}=0$), the Convergence scalar ($R_{c}$) coincides with the Curvature scalar ($\tilde{R}$) and $\tilde{R}$ can be treated as the Raychaudhuri scalar if we consider a congruence of time-like geodesics ($\nabla_{b}A^{b}=0$) which are hyper-surface orthogonal ($\omega=0$) and moving in a four dimensional space-time ($n=4$). As already mentioned in the work of \cite{Albareti:2014dxa},  $\tilde{R}$ can be treated as the mean curvature geometrically. Thus we attempt to find the solution of equation (\ref{eq8}) in FLRW background assuming some possible forms of $\tilde{R}$. Since the background is homogeneous, $\tilde{R}$ may be treated as a function of the scale factor $a$ say $\dfrac{\tilde{R}}{3}=G(a)$. Thus eq (\ref{eq8}) takes the form
	\begin{equation}
		\dfrac{d^{2}a}{dt^{2}}+G(a)a=0.
	\end{equation}
	Now we find the solution of the above differential equation considering the following cases.

\begin{figure}[h!]
		\includegraphics[height=7cm, width=7cm]{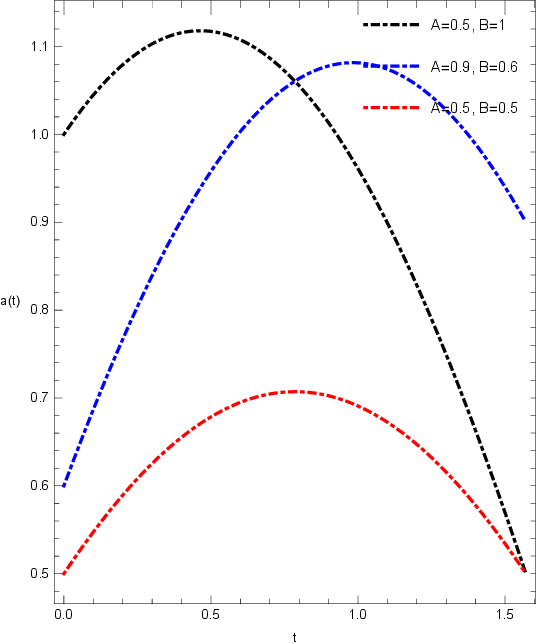}
		\centering \caption{$a(t)$ vs $t$ for constant positive mean curvature considering various parameters specified in the panel.}\label{f2}
	\end{figure}

	\textbf{Case-I:} $G(a)=l;~l>0$ i.e, we consider a positive constant mean curvature. The solution is given by 
	\begin{equation}
		a(t)=A~\sin(\sqrt{l}t)+~B~\cos(\sqrt{l}t).
	\end{equation} The variation of the scale factor $a(t)$ with cosmic time $t$ is shown graphically in FIG-(\ref{f2}) considering $l=1$ and different choices of $A$ and $B$.

\begin{figure}[h!]
		\includegraphics[height=7cm, width=7cm]{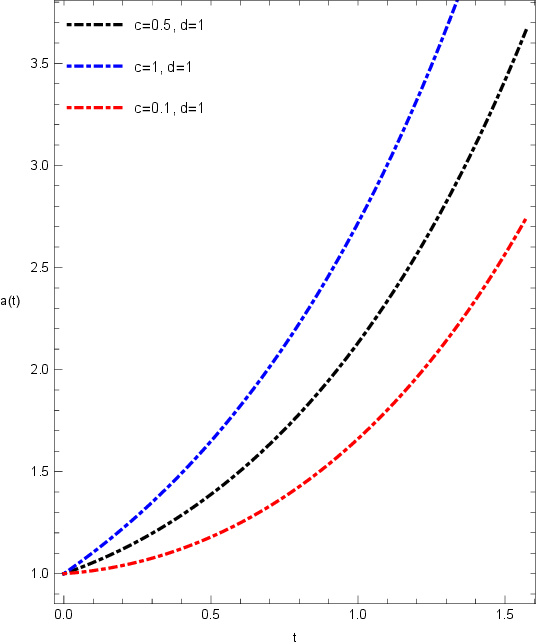}
		\centering \caption{$a(t)$ vs $t$ for constant negative mean curvature considering various parameters specified in the panel.}\label{f3}
	\end{figure}

	\textbf{Case-II:} $G(a)=-p~;p>0$ i.e, we consider constant negative mean curvature. In this case the solution is given by 
	\begin{equation}
		a(t)=c~\sinh(\sqrt{p}t)+d~\cosh(\sqrt{p}t).
	\end{equation} The graph of $a(t)$ vs $t$ has been shown in FIG-(\ref{f3}) for $p=1$ and different choices of the parameters involved.

\begin{figure}[h!]
		\includegraphics[height=7cm, width=7cm]{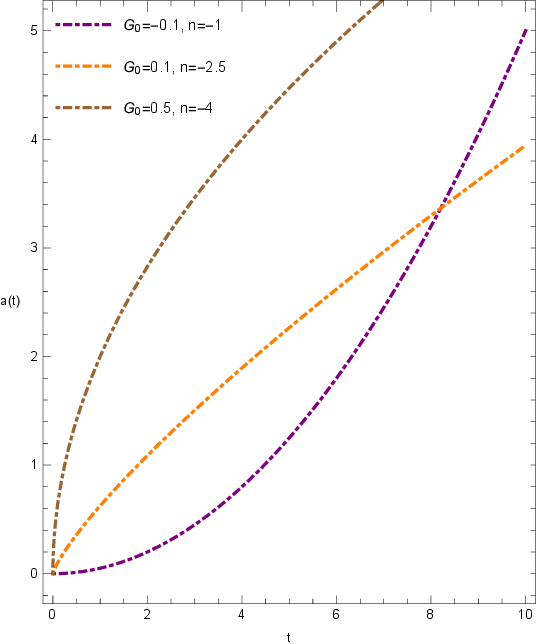}
		\centering \caption{$a(t)$ vs $t$ for variable mean curvature considering various parameters specified in the panel.}\label{f4}
	\end{figure}

	\textbf{Case-III:} $G(a)=G_{0}a^{n}$. In this case the solution assumes the form
	\begin{equation}
		a(t)=-\dfrac{G_{0}n^{2}}{2(n+2)}t^{-\frac{2}{n}}
	\end{equation} and graphically it is depicted in FIG- (\ref{f4}) for different choices of the parameters involved.
	\\ \\
	The above analysis shows that
	\begin{enumerate}
		\item In case of positive constant curvature we get bouncing scale factor (there is an epoch with respect to which there are two phases- expanding and contracting phase). In this case although the CC holds yet there is no curvature singularity. This is in agreement with the result of Landau and Lifschitz which reveals that CC or focusing alone can not imply singularity.
		\item In case of negative constant mean curvature, the CC is violated. The graph of $a(t)$ shows that there is no singularity. Thus in this case  avoidance of geodesic focusing or violation of the CC leads to possible avoidance of singularity.
		\item In case of variable mean curvature (positive definite) CC holds. This shows that at a strong curvature singularity, the gravitational tidal forces linked with the singularity are so strong that any particle trying to cross it must be crushed to zero size which is evident from FIG \ref{f4}.
	\end{enumerate}
	Thus, the above study shows that CC may or may not imply existence of singularity.
	
		Focusing Theorem for the flat model of universe is well known in literature. However if one includes curvature then what happens? Motivated by this question, in this section we try to rewrite the Focusing theorem for closed and open model of the universe, sign of curvature being positive and negative respectively. The first Friedmann equation with non-zero curvature term $K$ is given by
	\begin{equation}
		3\left(H^{2}+\dfrac{K}{a^{2}}\right)=\rho.\label{eq11}
	\end{equation}
	In FLRW background the expansion scalar is  $\Theta=3H$ where $H=\dfrac{\dot{a}}{a}$ is the Hubble parameter and $a(t)$ is the cosmic scale factor. Writing (\ref{eq11}) in terms of $\Theta$ we have
	\begin{equation}
		\dfrac{\Theta^{2}}{3}=\rho-\dfrac{3K}{a^{2}}.
	\end{equation} Using the RE in FLRW model ($\sigma=0$) considering hyper-surface orthogonal ($\omega=0$) congruence of time-like geodesic ($\nabla_{b}A^{b}=0$)  and $\tilde{R}=\frac{1}{2}(\rho+3p)$ one has
	\begin{equation}
		\dfrac{d\Theta}{dt}=\dfrac{3K}{a^{2}}-\dfrac{3}{2}(\rho+p)\label{eq13}
	\end{equation}
	Now we discuss the following cases:\\
	\textbf{Case-I : $K<0$} \\
	If $(\rho+p)\geq0$ then equation (\ref{eq13}) gives $\dfrac{d\Theta}{dt}<0$. Thus the expansion of the congruence decreases with time. In other words if we consider an open model of universe with matter satisfying the Null Energy Condition (NEC) then congruence may focus either in finite or in infinite time. This may lead to formation of congruence singularity. However if $(\rho+p)<0$ and $|\rho+p|>\dfrac{3K}{a^{2}}$ then $\dfrac{d\Theta}{dt}>0$. Therefore open universe with exotic matter can avoid focusing and hence singularity formation.\\ \\
	\textbf{Case II: $K>0$} \\ In this case $\dfrac{K}{a^{2}}>0$ and using equation (\ref{eq13}) one may find that if  $(\rho+p)>\dfrac{2K}{a^{2}}>0$ then $\dfrac{d\Theta}{dt}<0$. Therefore in a closed model of universe if matter satisfies the NEC with a non zero lower bound then expansion of the congruence decreases with time. However singularity may be avoided even with the assumption of NEC on matter provided $\dfrac{2K}{a^{2}}>(\rho+p)$. Thus a high positive curvature and matter content satisfying NEC is required in a closed universe for the possible avoidance of singularity. But in any case exotic matter always avoids the singularity.
	
	The cosmic rate of expansion is given by 
	\begin{equation}
		\dfrac{d\Theta}{dt}=-3H^{2}(1+q)
	\end{equation}
	Thus, $\Theta$ is decreasing for $q>-1$ and increasing for $q<-1$. So, negative expansion of geodesic is possible in the era of cosmic evolution characterized by $q>-1$ (non phantom era). On the other hand, the cosmic era determined by $q<-1$ (phantom era) has expanding flow of geodesic.
	\section{III. Raychaudhuri scalar and singularity} 
	\subsection{III.I Black-Hole} 
	In order to analyze Black-Hole singularity using the Raychaudhuri equation we consider the Schwarzschild metric
	\begin{equation}
		ds^{2}=-\left(1-\dfrac{2GM}{r}\right)dt^{2}+\left(1-\dfrac{2GM}{r}\right)^{-1}dr^{2}+r^{2}d\Omega^{2}\label{eq*1}.
	\end{equation} Schwarzschild metric is a vacuum solution of the Einstein's field equations and in general relativity a vacuum solution is a Lorentzian manifold whose Einstein tensor (and hence the Ricci tensor $R_{ab}$) vanishes identically and thus $\tilde{R}=0$. Therefore the Convergence scalar $R_{c}=2\sigma^{2}$. By the dynamics of classical Schwarzschild metric \cite{Blanchette:2021jcw} and using the definition of anisotropy scalar ($\sigma^{2}$) one has
	\begin{equation}
		R_{c}=2\sigma^{2}=\dfrac{4}{3r^{3}}\dfrac{(3GM-r)^{2}}{(2GM-r)}.\nonumber
	\end{equation} Based on the above expression of $R_{c}$ one has the following findings:\\ \\
	$~~~~~$(i) $R_{c}>0$ only when  $r<2GM$ . So singularity may be possible only when  $r<2GM$. Further as $r\rightarrow0$  a stage will come when $R_{c}$ will predominate over $\omega^{2}$ and $\nabla_{c}A^{c}$ so that $R_{c}-2\omega^{2}-\nabla_{c}A^{c}\geq0$ (Convergence Condition) and this hints that convergence and hence formation of singularity may occur.\\ 
	
	(ii) For $r>2GM$, $R_{c}<0$. So singularity is not possible in the region $r>2GM$. Therefore $r=2GM$ acts a boundary to distinguish the regions with and without singularity. Here $r=2GM$ is nothing but the Event-Horizon.\\
	
	(iii) $r=0$ and $r=2GM$ are the points where $R_{c}$ diverges/ blows off. These are therefore identified as the points of singularity. However singularity at $r=2GM$ is a co-ordinate singularity that arises due to bad choice of coordinates and can be removed in Eddington-Finkelstein co-ordinates. On the other hand $r=0$ is the physical singularity or the Black-Hole singularity whose existence is inevitable from the point of view of the RE. Further from the Harmonic oscillator approach, we have $\omega_{0}^{2}=R_{c}$, the frequency of the oscillator. In the expression for $R_{c}$ if $r=o(\epsilon)$ then as $r\rightarrow0$, $\omega_{0}^{2}$ varies largely with $o(\epsilon)^{3}$ as compared to $r\rightarrow 2GM$ where the frequency varies with $o(\epsilon)$. This is in agreement with the prediction of General Relativity that the space-time near a cosmological singularity undergoes an infinite number of oscillations between different Kasner epochs with rapid transitions between
	them \cite{Casalderrey-Solana:2023zlg},\cite{Belinsky:1970ew}.

	Therefore, RE tells us that the space-time may	develop singularity due to presence of anisotropy and Black-Hole singularity is an example of
	such a case. Thus we essentially show the existence of black-hole singularity using the classical RE. However, quantum resolution of black-hole singularity is not attempted in the present work. It can be found in \cite{Das:2013oda}, \cite{Ashtekar:2008ay}, \cite{Blanchette:2020kkk}.
	\subsection{ III.II Cosmology}
	$\tilde{R}$ has a nice interpretation in cosmology regarding the convergence of the congruence of time-like geodesics (also known as focusing of geodesics) and in the avoidance of singularity as follows:\\
	In Einstein gravity or in usual modified gravity the field equations for gravity can be written as
	\begin{equation}
		G_{\mu\nu}=\kappa~ T_{\mu\nu}^{'}~,
	\end{equation}
	where $T_{\mu\nu}^{'}=T_{\mu\nu}$ is the usual energy-momentum tensor for the matter field in Einstein gravity while $T_{\mu\nu}^{'}=T_{\mu\nu}+T_{\mu\nu}^{(e)}$ for most of the modified gravity theories with $T_{\mu\nu}^{(e)}$ containing the extra geometric/physical terms in the field equations. Thus the Raychaudhuri scalar/ Curvature scalar $\tilde{R}$ takes the following form in terms of the energy-momentum tensor or/and the effective energy-momentum tensor as:
	\begin{equation}
		\tilde{R} =\kappa~(T_{\mu\nu}-\frac{1}{2}Tg_{\mu\nu})u^{\mu} u^{\nu}=\dfrac{1}{2}\left(\rho+3p\right),~Einstein~ Gravity~~\label{eq6}
	\end{equation}
	\begin{eqnarray}
		\small	= \kappa~\left[(T_{\mu\nu}-\frac{1}{2}Tg_{\mu\nu})u^{\mu} u^{\nu}+(T^{(e)}_{\mu\nu}-\frac{1}{2}T^{(e)}g_{\mu\nu})u^{\mu} u^{\nu}\right]\nonumber\\
		\small	=\dfrac{1}{2}\left(\rho+3p\right)+\dfrac{1}{2}\left(\rho^{(e)}+3p^{(e)}\right),~Modified ~Gravity~~~\label{eq7***}
	\end{eqnarray}
	Here we show a three fold interpretation of $\tilde{R}$ which lead us to the same conclusion regarding the convergence and avoidance of singularity.
	\begin{enumerate}
		\item 
		In FLRW space-time the (effective) Einstein field equations are
		\begin{eqnarray}
			\scriptsize 	3H^{2}=\kappa \rho,~~2\dot{H}=-\kappa(\rho+p) \label{eq6**}\\
			\scriptsize	3H^{2}=\kappa(\rho+\rho_{e}),~~2\dot{H}=-\kappa[(\rho+p)+(\rho_{e}+p_{e})]\label{eq7**}
		\end{eqnarray} where equation (\ref{eq6**}) is for Einstein gravity, (\ref{eq7**}) corresponds to Modified gravity and `.' is differentiation w.r.t cosmic time $t$.\\
		So the deceleration parameter $q=-\left(1+\dfrac{\dot{H}}{H^{2}}\right)$ takes the form:
		\begin{eqnarray}
			q=\dfrac{\rho+3p}{2\rho}, ~Einstein~ gravity.\nonumber\\
			~~~~q=\dfrac{(\rho+3p)+(\rho_{e}+3p_{e})}{2(\rho+\rho_{e})},~ Modified ~gravity.\nonumber
		\end{eqnarray}
		Hence 
		\begin{eqnarray}
			\tilde{R}=q\rho~, Einstein~gravity\nonumber
			\\ 
			\tilde{R}=q(\rho+\rho_{e})~, Modified~gravity\nonumber
		\end{eqnarray} One finds that $\tilde{R}=\dfrac{3qH^{2}}{\kappa}$ for both the cases. Now for convergence $\tilde{R}>0$ so one may conclude that convergence will occur during the evolution of the universe if $q>0$. One may interpret $q$ as the convergence scalar. Thus formation of singularity is not possible both in early inflationary era and the present late time accelerated era of evolution, while the matter dominated era of evolution favors the convergence. This formulation brings out an inherent feature of the deceleration parameter as convergence scalar.
		\item From eq (\ref{eq6}) (i.e in Einstein gravity) if we write $\rho=\rho_{1}+\rho_{2}$ and $p=p_{1}+p_{2}$ then we find that  $\rho_{2}+3p_{2}<0$ and
		$|\rho_{2}+3p_{2}|> \rho_{1}+3p_{1}$  yield $\tilde{R}<0$. In other words $(\rho_{2},~p_{2})$ corresponds to the energy density and thermodynamic pressure of dark energy. Hence dominance of dark energy over normal matter ( having density and pressure ($\rho_{1},~p_{1}$)) does not allow Focusing Theorem to hold. Therefore the era dominated by dark energy namely the inflationary era and the present accelerated era of expansion are against the formation of singularity.
		\item From eq (\ref{eq7***}) (i.e. in modified gravity), to make $\tilde{R}<0$ we need $\rho^{(e)}+3p^{(e)}<0$ and $|\rho^{(e)}+3p^{(e)}|> \rho+3p$. Again it hints that $(\rho^{(e)},p^{(e)})$ corresponds to the density and pressure of dark energy and hence we arrive at the same conclusion as in the former cases.
	\end{enumerate}
	Raychaudhuri equation in terms of cosmic parameters for hyper-surface orthogonal congruence of time-like geodesic in isotropic background can be written as 
	\begin{equation}
		\dot{H}=-(1+q)H^{2}\label{eq24***}
	\end{equation}
	Thus RE can be written in two ways: one in terms of geometric or kinematic variables and the other in terms of cosmic parameters given by the above equation. We call the former as geometric form of RE and the later as cosmological form of RE. In this context, one may observe a nice analogy in these two forms. In geometric form $\tilde{R}$ or the Raychaudhuri scalar plays the role in convergence or focusing while in cosmological form $q$, the deceleration parameter does it. Moreover, the effect of gravity theory comes through $\tilde{R}$ in the geometric form while in the cosmological form $q$ carries the effect of gravity. Therefore $q$ has a similar role as $\tilde{R}$. Further the geometric form of the RE can be converted to the evolution equation of a Harmonic Oscillator using a suitable transformation of variable.  One may refer to \cite{Chakraborty:2023ork} for details of Harmonic oscillator equation derived from the geometric form of RE. The paper \cite{Chakraborty:2023ork} gives a transformation under which the first order RE can be converted to the Harmonic oscillator equation and it can be shown that the convergence condition, avoidance of singularity everything are related to the time varying frequency of the oscillator. Following this approach, in this paper we attempt to show the Harmonic oscillator equation from the cosmological form of the RE. For this, we consider the cosmological RE. Using,
	\begin{equation}
		H=\dfrac{\dot{Y}}{Y}
	\end{equation} the first order cosmological form of the RE can be converted to a second order differential equation analogous to the evolution equation of a classical real harmonic oscillator as 
	\begin{equation}
		\ddot{Y}+qH^{2}Y=0.
	\end{equation} The frequency of the real harmonic oscillator is given by $W^{2}=qH^{2}$. Thus, formation of a real Harmonic oscillator is characterized by positive $q$ and hence possible only in the matter-dominated era and not in early inflationary era or at the present era.  As $W^{2}$ should rapidly change near singularity, '$q$' and $H$ should have rapid fluctuation i.e, $H$ and $\dot{H}$ will
	have highly oscillatory behavior. Cosmological RE predicts whether we will encounter singularity or not in the course of evolution.
	\section{IV. Quantum mitigation of classical singularity}
	\subsection{IV.I Lagrangian and Hamiltonian formulation of RE: A general quantum description}
	In this section we aim to find the evolution of a quantized time-like geodesic congruence or the quantized RE. To do so we consider  \cite{Alsaleh:2017ozf} 
	\begin{equation}
		x(\tau)=\sqrt{q}
	\end{equation} so that
	\begin{equation} \dfrac{1}{\sqrt{q}}\dfrac{d\sqrt{q}}{d\tau}=\Theta
	\end{equation} i.e
	\begin{equation}
		\dfrac{dx}{d\tau}=x\Theta,
	\end{equation}
	where $q=$ det ($q_{ab})$ and ($q_{ab}$ is the metric on the space-like hyper-surface). Essentially $x$ is related to the volume
	of the hyper-surface and $x = 0$ identifies the singularity. As a result the first order non-linear RE becomes a second order non-linear ordinary differential equation as 
	\begin{equation}\label{eq5}
		x\dfrac{d^{2}x}{d\tau^{2}}+\left(\dfrac{dx}{d\tau}\right)^{2}\left(\dfrac{1}{n}-1\right)+x^{2}~R_{c}=0.
	\end{equation}  The above second order differential equation (\ref{eq5}) possesses a first integral of the form \cite{Chakraborty:2023ork}
	\begin{equation}\label{eq4}
		\left(\dfrac{dx}{d\tau}\right)^{2}=z_{0}~ x^{2(1-\frac{1}{n})} -2~x^{2(1-\frac{1}{n})}\int x^{(\frac{2}{n}-1)}~R_{c}~dx
	\end{equation}
	i.e.
	\begin{equation}
		\Theta^{2}=z_{0}~ x^{-\frac{2}{n}}-2 x^{-\frac{2}{n}}\int x^{(\frac{2}{n}-1)}~R_{c}~dx
	\end{equation} with
	$z_{0}$, the constant of integration. It is interesting to note that $z_{0}$ is not merely a constant of integration but is related to the curvature (geometry) of the space-time as the above first integral can be identified with the 1st Friedmann equation ($3H^{2}+\dfrac{\kappa}{a^{2}}=\rho$) in FLRW model (where $\Theta=3H$) with $\kappa=-\dfrac{z_{0}}{3}$.\\
	Moreover the RE (\ref{eq2}) can be simplified to a great extent if we consider a family of time-like geodesics (i.e $A^{b}=0$) orthogonal to a space-like hyper-surfaces (i.e $\omega_{ab}=0$) as
	\begin{equation}
		\dfrac{d\Theta}{d\tau}=-\dfrac{\Theta^{2}}{3}-R_{c}.
	\end{equation}
	As the RE can be written in the form of a second order differential equation (\ref{eq5}), so it is a natural question whether there exists a Lagrangian for which the Euler-Lagrange equation corresponds to the RE (i.e eq. (\ref{eq5})). Also it is to be noted that for quantum description one needs the Hamiltonian formulation (and hence a Lagrangian formulation). Usually, a second order differential equation will be the Euler-Lagrange equation corresponding to a Lagrangian if it satisfies the Helmholtz conditions \cite{Crampin:2010}, \cite{Nigam:2016}. Therefore if we write,
	\begin{equation}
		\mu=x\frac{d^{2}x}{d\tau^{2}}+\left(\frac{dx}{d\tau}\right)^{2}\left(\frac{1}{n}-1\right)+x^{2}R_{c}
	\end{equation} then it shows that $\mu$ fails to satisfy all the Helmholtz conditions. However
	\begin{equation} \tilde{\mu}=x^{(\frac{2}{n}-3)}\mu
	\end{equation} satisfies all the Helmholtz conditions, provided $R_{c}$ is a function of $x$ alone say $g(x)$. Thus one may choose the Lagrangian as 
	\begin{equation}
		\mathcal{L}=\dfrac{1}{2}x^{2(\frac{1}{n}-1)}\left(\frac{dx}{d\tau}\right)^2-V[x]
	\end{equation} with 
	\begin{equation}
		\dfrac{dV[x]}{dx}= x^{(\frac{2}{n}-1)}g(x).
	\end{equation} The conjugate momentum to the variable $x$ is $\Pi_{x}=x^{2(\frac{1}{n}-1)}\frac{dx}{d\tau}$. Hence the Hamiltonian of the system is
	\begin{equation} \mathcal{H}=\dfrac{\Pi_{x}^{2}}{2x^{2(\frac{1}{n}-1)}}+V[x].
	\end{equation} It is interesting to note that one of the Hamilton's equation of motion gives the RE and the other yields the definition of momentum. Further one finds that $R_{c}=g(x)$ is nothing but the gradient of the potential $V$, and hence $R_{c}>(<)~0$ implies force is attractive (repulsive) in nature. It shows that convergence will occur (i.e. $R_{c}>0$) if the matter is attractive. This is the reason why RE can be called as the fundamental equation of gravitational attraction.\\

	\textbf{IV.I.I Wheeler-DeWitt formalism: Canonical Approach}
	
	For canonical quantization, $x$ and $\Pi_{x}$ are considered as operators acting on the state vector $\Psi(x,\tau)$ of the geometric flow. In $x$- representation, the operators assume the form $\tilde{x}\rightarrow x$ and $\tilde{\Pi_{x}}\rightarrow -i\hbar \dfrac{\partial}{\partial x}$ so that $[\tilde{x}, \tilde{\Pi_{x}}]=i\hbar$ and the operator form of the Hamiltonian is
	\begin{equation}
		\tilde{\mathcal{H}}=-\dfrac{\hbar^{2}}{2} x^{2(1-\frac{1}{n})} \dfrac{d^{2}}{dx^{2}}+V[x].
	\end{equation} In the context of cosmology, there is notion of Hamiltonian constraint and operator version of it acting on the wave function of the universe ($\Psi$) gives $\tilde{\mathcal{H}}\Psi=0$. This is known as the Wheeler-Dewitt (WD) equation which explicitly takes the form
	\begin{equation}\label{eq7*} \dfrac{d^{2}\Psi}{dx^{2}}-\dfrac{2}{\hbar^{2}}x^{2(\frac{1}{n}-1)}V[x]\Psi(x)=0.
	\end{equation} The problem of non-unitary evolution can be resolved by proper operator ordering in the first term of the Hamiltonian. The above WD equation has no time evolution and so Hamiltonian annihilates the wave function. Thus, here the problem of factor ordering is not related to the unitary evolution. \\If one considers the following operator ordering 
	\begin{equation}
		\tilde{\mathcal{H}}=-\dfrac{\hbar^{2}}{2} x^{(1-\frac{1}{n})}\dfrac{d}{dx} x^{(1-\frac{1}{n})}\dfrac{d}{dx}+V[x],
	\end{equation} then by choosing $v=nx^{\frac{1}{n}}$ the WD equation is transformed as 
	\begin{equation}\label{eq7}
		\left(-\dfrac{\hbar^{2}}{2} \dfrac{d^{2}}{dv^{2}}+V[v]\right)\Psi(v)=0,
	\end{equation} with symmetric norm as $|\Psi|^{2}=\int_{0}^{\infty} dv \Psi^{*} \Psi$, provided the integral exists and finite.
	Utility of this quantization is as follows:
	\begin{itemize} 
	\item The WD equation (\ref{eq7}) can be interpreted as time-independent Schrödinger equation of a point particle of unit mass moving along $v$ direction in a potential field $V(v)$ and it has zero eigen  value of the Hamiltonian and the wave function of the universe is identified as the energy eigen function. 
		\item 
		$V [v]$ is the potential corresponding to
		the dynamical system representing the congruence and has to be constructed using the
		gravitational field equations. In case of homogeneous cosmology $V[v]=V(v)$. For inhomogeneous cosmologies, the method will not work,
		as $V$ will remain a functional. 
		\item For any modified gravity theory constructed in the background of homogeneous (isotropic/anisotropic) space-time one may find the classical potential $V$. Further, if one can solve the WD equation and find $|\Psi|^{2}$  then it can be used as a quantity proportional to the probability measure on the minisuperspace. If $|\Psi|^{2}=0$ at classical singularity (characterized by $a\rightarrow0$) then singularity is avoided otherwise the singularity still persists in the quantum description.  This boundary condition is purported to circumvent the initial big-bang singularity and is known as DeWitt conjecture \cite{DeWitt:1967yk}. However, later it has been shown in \cite{Gotay:1983kvm} that the DeWitt boundary condition does not actually play a role to avoid singularity (though there are some controversies regarding its fundamental relevance). More generally, quantum states of the singularity can be examined by calculating the expectation values of the observables which vanish at the singularity classically. For details one may refer to \cite{Gotay:1983kvm}, \cite{Gotay:1980xk}, \cite{Lund:1973zz}. Mathematically, a quantum state $|A>$ is considered as singular if every operator $\Omega$ corresponding to a classical observable should have a vanishing expectation value i.e, $<A|\Omega|A>=0$ at the singularity \cite{Jalalzadeh:2022dlj}.
	
		\item One may note that the existence (or non existence) of singularity is not a generic one, it depends on the gravity theory under consideration through the convergence scalar $R_{c}$ via the classical potential $V$. 
		\item This quantization is expected to find application in the investigation of the
			singularities in the quantum regime for a collapse of homogeneous systems, such as the
			Datt-Oppenheimer-Snyder collapse \cite{Oppenheimer:1939ue}, \cite{Ziaie:2024saq}.
		\item Thus the  quantization of the geometric flow of a congruence of  classical geodesic gives possibilities of avoidance of singularity in homogeneous cosmologies.
	\end{itemize}
	Application-I:
	In cosmology, there are two phenomenological choices for the potential namely (i) $V=V_{0}v^{n}$ and (ii) $V=V_{0}\exp(-\lambda v)$. The justification for such a power law form is that it satisfies the sufficiently steep condition namely $\Gamma=\dfrac{V''V}{(V')^{2}}\geq 1$. As a result, the scalar field rolls down such a potential and favors a common evolutionary path for a wide range of initial conditions \cite{Ratra:1987rm}, \cite{Chakraborty:2024nmg}, \cite{vs}. The choice of exponential potential corresponds to an extreme example of quintessence \cite{Ratra:1987rm}, \cite{Chakraborty:2024nmg}, \cite{vs}. Exponential potential will remain subdominant if it was so initially. Further, nucleosynthesis constraints the energy density of the quintessence field to be smaller than the associated background energy density at the early era. 
	
	For the choice (i), the WD equation reduces to
	\begin{equation}
		\dfrac{d^{2}\Psi(v)}{dv^{2}}-\dfrac{2V_{0}}{\hbar^{2}}v^{n}\Psi(v)=0
	\end{equation} which has a solution of the form
	\begin{equation}	\Psi(v)=\\(2V_{0})^{\frac{1}{4+2n}}(\hbar(n+2))^{\frac{-1}{n+2}}\sqrt{v}\mathcal{L}
	\end{equation} where $\mathcal{L}=L_{1}+L_{2}$. $L_{1}$ and $L_{2}$ are respectively given by
	\begin{equation} 
		L_{1}=c_{1}I_{\frac{-1}{n+2}}\left(\frac{2\sqrt{2V_{0}}v^{1+\frac{n}{2}}}{(n+2)\hbar}\right)\Gamma \left(\frac{1+n}{2+n}\right)\nonumber
	\end{equation} and,
	\begin{equation}
		L_{2}=c_{2}(-1)^{\frac{1}{2+n}}I_{\frac{1}{n+2}}\left(\frac{2\sqrt{2V_{0}}v^{1+\frac{n}{2}}}{(n+2)\hbar}\right)\Gamma \left(\frac{3+n}{2+n}\right)\nonumber
	\end{equation}
	On the other hand choice (ii) yields the WD equation and its solution as follows
	\begin{equation}
		\dfrac{d^{2}\Psi(v)}{dv^{2}}-\dfrac{2V_{0}}{\hbar^{2}}\exp(-\lambda v)\Psi(v)=0
	\end{equation} and 
	\begin{equation}
		\small	\Psi(v)=d_{1} I_{0}\left(\dfrac{2\sqrt{2V_{0}}\sqrt{\exp(-\lambda v)}}{ \lambda \hbar}\right)+2 d_{2}K_{0}\left(\dfrac{2\sqrt{2V_{0}}\sqrt{\exp(-\lambda v)}}{\lambda \hbar}\right)
	\end{equation} respectively. Here $c_{1}$, $c_{2}$, $d_{1}$ and $d_{2}$ are arbitrary constants. $I_{\nu}(z)$ and $K_{\nu}(z)$ stand for Bessel $I(\nu,z)$ and $K(\nu,z)$ functions respectively. Thus the above formulation says that, for power law potential $|\Psi|^{2}=0$ at classical singularity ($v=0$) while $|\Psi|^{2}\neq0$ at $v=0$ in case of  choice (ii) of the potential. Thus for exponential potential, there is always a non zero finite probability of having a singularity while in the power law case the probability of having zero volume is zero.\\

	\textbf{IV.I.II Bohmian Formalism: Causal Approach}\\

\begin{figure}[h!]
		\includegraphics[height=6cm,width=7cm]{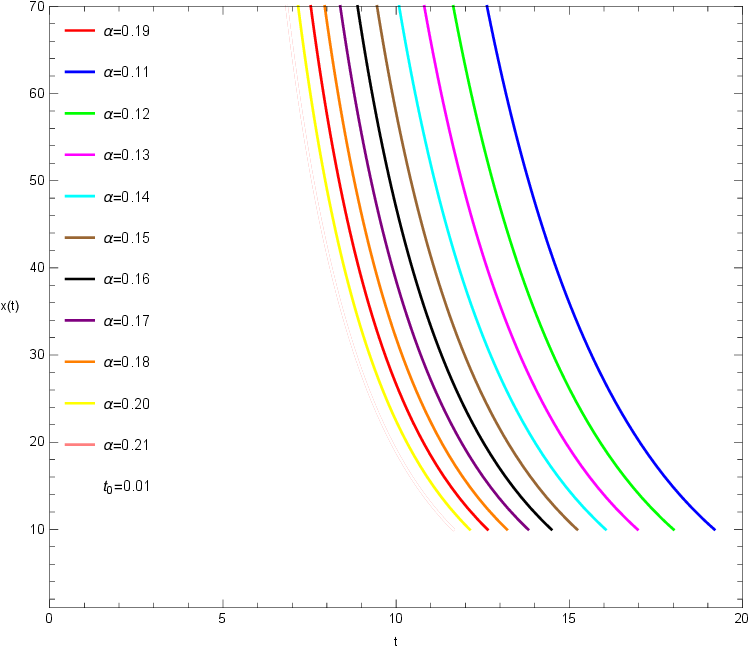}
		\centering \caption{Bohmian trajectories for non zero quantum potential.}\label{f5}
	\end{figure}

	 Application-II:
	Unlike the previous section where we quantized the geodesic flow in this section we replace them by quantum Bohmian trajectories.  A detailed overview of these trajectories can be found in the works \cite{Das:2013oda}, \cite{Ali:2014qla}, \cite{Chakraborty:2023voy}.  Actually, Bohmian trajectories are the quantum trajectories derived from Hamilton Jacobi equation which is an equation obtained by using an ansatz for the wave function into the WD equation. Unlike the case of classical geodesics, here there is a notion of quantum potential that carries the quantum effect. For the present system we choose an ansatz for the wave function as \cite{Chakraborty:2001za}
	\begin{equation}\label{eq90*}
		\Psi(x)= A(x)~\exp\left(\frac{i}{\hbar}~S(x)\right).
	\end{equation}
	Using this ansatz into the WD - equation (\ref{eq7*}) one gets the Hamilton-Jacobi equation as 
	\begin{equation}
		\dfrac{-1}{2~x^{2(\frac{1}{n}-1)}}~\left(\dfrac{dS}{dx}\right)^{2} + V_{Q} + V(x)=0,
	\end{equation}
	where $V_{Q}$, the quantum potential has the expression as
	\begin{equation}
		V_{Q}=\dfrac{1}{2 A(x)~x^{2(\frac{1}{n}-1)}} \dfrac{d^{2}A(x)}{dx^{2}}.
	\end{equation}  Thus the Hamilton-Jacobi function $S$ is given by 
	\begin{equation}\label{eq93}
		S=s_{0} \pm \int\left(\dfrac{1}{A(x)}\dfrac{d^{2}A(x)}{dx^{2}}+ 2x^{2(\frac{1}{n}-1)}\right)^{\frac{1}{2}}~dx,
	\end{equation} where $s_{0}$ is the constant of integration.\\
	It may be noted that the trajectories $x(t)$ due to causal interpretation should be real, independent of any observation and are classified by the above H-J equation. In fact, the quantum trajectories i.e the Bohmian trajectories are first order differential equations characterized by the equivalence of the usual definition of momentum with that from the Hamilton-Jacobi function $S$ as 
	\begin{equation}
		\dfrac{dS(x)}{dx}=-2x^{2(\frac{1}{n}-1)}\dfrac{dx}{dt}
	\end{equation}or,
	\begin{equation}\label{eq96}
		2~\int \dfrac{x^{2(\frac{1}{n}-1)}~dx}{\left(\dfrac{1}{A(x)}~\dfrac{d^{2}A(x)}{dx^{2}}+2x^{2(\frac{1}{n}-1)}\right)^{\frac{1}{2}}}=\mp (t-t_{0}).
	\end{equation}
For cosmological description in FLRW model in quantum era we consider scalar field with self interacting potential as the matter content . Then one may have power law form of the scale factor i.e, 
	\begin{equation}
		a(t)=a_{0}t^{K},
	\end{equation} where $a_{0},~K$ are constants with $\phi=\phi_{0}\ln a$ and $V(\phi)=V_{0}\exp(\mu\phi)$.
	Now we consider the following cases:\\
	\textbf{Case-A}
	$A(x)=A_{0}$, a constant then from (\ref{eq93}) one has $
	S=s_{0}\pm \sqrt{2}~n~x^{\frac{1}{n}}
	$ and the quantum trajectory is described as,
	\begin{equation}
		\sqrt{2}~n~x^{\frac{1}{n}}=\pm (t-t_{0}).
	\end{equation}
	Here the quantum potential ($V_{Q}$) is zero and the H-J equation coincides with the classical one. Thus Bohmian trajectory corresponds to classical power law form of expansion and it can't avoid the initial big-bang singularity. \\ \\
	\textbf{Case-B} $A(x)=x^{L}$, $L\in(0,1)$ and substituting this in equation (\ref{eq96}) one gets the quantum trajectories as,
	\begin{equation}
		x(t)=\left[\frac{(t-t_{0})^{2}}{2n^{2}}-\frac{L(L-1)}{2}\right]^{\frac{n}{2}}.
	\end{equation}
	In this choice the quantum potential has non zero value and the Bohmian trajectories can avoid the initial classical singularity.\\ \\
	\textbf{Case-C} $A(x)=A_{0}\exp(-\alpha x),~\alpha\neq0$.
	The quantum trajectory is described by
	\begin{equation}
		\dfrac{6}{\alpha x^{\frac{1}{3}}}2^{F_{1}}\left(\dfrac{1}{4},\dfrac{1}{2},\dfrac{5}{4},-\dfrac{2}{\alpha^{2}x^{\frac{4}{3}}}\right)=(t-t_{0}).
	\end{equation}  where $2^{F1}$ is the Gauss- Hypergeometric function. Clearly the quantum trajectory is a one parameter family of curves which never pass through classical singularity. The trajectories for the last case is shown graphically in FIG-(\ref{f5}).
	
	One may note that formulation of Bohmian trajectories is unlike the Wheeler-DeWitt quantization which involves the classical potential corresponding to the congruence. This classical potential has to be constructed using the field equations via the Raychaudhuri scalar ($\tilde{R}$) and employing it in the WD equation one can proceed to singularity analysis. Bohmian formalism studies the characteristics of quantum Bohmian trajectories near the initial big-bang singularity. Classical geodesics are destined to focus at the big-bang singularity but bohmian trajectories may defocus as and when the quantum potential is included in Hamilton Jacobi equation. We have recently applied the two techniques in some homogeneous and isotropic $f(T)$ gravity models in \cite{Chakraborty:2023yyz} and in mitigating Black-Hole singularity in \cite{Chakraborty:2024dpx}.
	
	\section{V. Conclusion and Discussion}To conclude, in this letter we have explored the classical and quantum description of singularity via the celebrated RE. A suitable transformation identifies the RE to the evolution equation of a classical Harmonic Oscillator with varying frequency and it is found that singularity is associated with this frequency. Further using the geometrical interpretation of the Raychaudhuri scalar as mean curvature we have found cosmological solution mainly the cosmic scale factor $a(t)$, for three different choices. This leads to the result that Convergence Condition (CC) may or may not lead to singularity formation. A three fold cosmological interpretation of the Raychaudhuri scalar has been given in the context of convergence and avoidance of singularity, which lead us to the same conclusion. By a suitable transformation the first order non-linear RE can be converted to a second order differential equation and a first integral of it has been found to be an analytic solution of the RE. Further this first integral matches with the first Friedmann equation in Einstein gravity.  Moreover Black-Hole singularity has been investigated using the RE. Focusing theorem for flat model is well known in literature. However in the present work we have restated the Focusing theorem for non zero curvature. Thus the existence of cosmological or black hole singularity is successfully described by the RE.
	Further the present work also shows quantum resolution of the initial big-bang singularity in mainly two ways. Firstly, the Lagrangian and the Hamiltonian formulation for the RE has been carried out for canonical quantization. Further we have argued that the solution of the WD equation plays a pivotal role for singularity analysis in quantum regime. This is an attempt to quantize the geodesic flow and avoid singularity in homogeneous cosmology. This formulation is expected to find application in gravitational collapse of homogeneous systems. In the letter, we have essentially shown the ``existence" of black hole and cosmological singularity via the celebrated Raychaudhuri equation. However, we have attempted the quantum resolution of only the cosmological singularity namely, the initial big-bang singularity (by studying the nature of Bohmian trajectories near the classical singularity) and not the black-hole singularity. This is because, we have studied the evolution of a quantized time-like geodesic congruence (not null geodesic) in the present work. In principle, there is no basic difference between the use of these two kinds of geodesics. However, in the context of cosmology as the evolution is characterized by time variation so time-like geodesic is important where proper time or the cosmic time is being used. Similarly, if one carries out the quantization of a congruence of null geodesic and study their evolution w.r.t an affine parameter, then that should find an immediate application in black hole physics. One may interpret an affine parameter to be the null analog of proper time. Resolution of Black-hole singularity has been attempted recently in \cite{Chakraborty:2024dpx}. We have carried out the quantum formulation via the Lagrangian and Hamiltonian formulation of RE using a transformation related to the metric scalar of the hyper-surface. Next we invoke the notion of a Hamiltonian constraint and operator version of it acting on the wave function of the universe, known as Wheeler DeWitt (WD) equation. The potential involved in the WD equation is related to the convergence scalar $R_{c}=2\sigma^{2}+\tilde{R}$ where $\tilde{R}$ is the Raychaudhuri scalar. For any modified gravity theory in the homogeneous background, it is expected to obtain the potential and solve the WD equation. Norm square of the solution of WD equation (wave function of the universe) is then examined at the classical singularity $a\rightarrow0$ ($a$ is the cosmic scale factor).
	It is inferred that if $|\Psi|^{2}$ is zero as $a\rightarrow0$, then the probability of having a classical singularity is zero (DeWitt Conjecture). In other words, the classical singularity is avoided in quantum description. If $|\Psi|^{2}$ is non-zero, then there is a finite probability of having zero volume or singularity. Thus quantum description fails to mitigate the classical singularity.  However, we have another way of examining the classical singularity by examining the expectation value of the observable (as discussed earlier). Here, the only observable that we have is the energy operator or the Hamiltonian operator. As $\Psi$, obtained by solving the WD equation satisfies $\tilde{H}|\Psi>=0$, so the expectation value $<\Psi|\tilde{H}|\Psi>$ is always zero. Hence, we can not have any conclusion from the expectation value at the singularity.
	 While on the other hand, Bohmian formalism makes use of the Hamilton Jacobi equation. Using it we study the nature of Bohmian trajectories (quantum analogue of classical geodesics) near the classical singularity. To speak lucidly, whether they pass through the classical singularity or not is examined. The difference from WD formalism is that in Bohmian formalism quantum potential is taken into consideration along with the classical potential to study its effect in focusing/defocusing.  Replacement of classical geodesics by quantum Bohmian trajectories  obviate the Big Bang singularity in the presence of non-zero quantum potential. Therefore the present work studies the existence (both black-hole (via Harmonic oscillator approach of RE) as well as cosmological singularity (via the focusing theorem)) and resolution of the gravitational singularities using the Raychaudhuri equation.
	\section*{Acknowledgement}
	The authors are thankful to the editor and anonymous reviewers for their valuable and insightful comments that increased the quality and visibility of the letter. M.C thanks University Grants Commission (UGC) for providing the Senior Research Fellowship (ID:211610035684/JOINT CSIR-UGC NET JUNE-2021). S.C. thanks FIST program of DST, Department of Mathematics, JU (SR/FST/MS-II/2021/101(C)).
	

\begin{thebibliography}{50}
		\bibitem{raychaudhuri}
		Raychaudhuri, Amal K. "Theoretical cosmology." Clarendon  Press , Oxford Studies in Physics (1979).
		\bibitem{Raychaudhuri:1953yv}
		A.~Raychaudhuri,
		``Relativistic cosmology. 1.,''
		Phys. Rev. \textbf{98}, 1123-1126 (1955).
		\bibitem{Dadhich:2005qr}
		N.~Dadhich,
		``Derivation of the Raychaudhuri equation,''
		[arXiv:gr-qc/0511123 [gr-qc]].
		\bibitem{Dadhich:2007pi}
		N.~Dadhich,
		Pramana \textbf{69}, 23-30 (2007)
		\bibitem{Ehlers:2006aa}
		J.~Ehlers,
		Int. J. Mod. Phys. D \textbf{15}, 1573-1580 (2006).
		\bibitem{Kar:2008zz}
		S.~Kar,
		Resonance J. Sci. Educ. \textbf{13}, 319-333 (2008).
		\bibitem{Kar:2006ms}
		S.~Kar and S.~SenGupta,
		Pramana \textbf{69}, 49 (2007).
		\bibitem{Chakraborty:2023rgb}
		M.~Chakraborty and S.~Chakraborty,
		Annals Phys. \textbf{460}, 169577 (2024)
		\bibitem{Penrose:1964wq}
		R.~Penrose,
		Phys. Rev. Lett. \textbf{14}, 57-59 (1965).
		\bibitem{Hawking:1970zqf}
		S.~W.~Hawking and R.~Penrose,
		Proc. Roy. Soc. Lond. A \textbf{314}, 529-548 (1970).
		\bibitem{Hawking:1973uf}
		S.~W.~Hawking and G.~F.~R.~Ellis,
		Cambridge University Press, (2011).
		\bibitem{DeWitt:1967yk}
		B.~S.~DeWitt,
		Phys. Rev. \textbf{160}, 1113-1148 (1967)
		\bibitem{Hawking:1978jz}
		S.~W.~Hawking,
		Phys. Rev. D \textbf{18}, 1747-1753 (1978)
		\bibitem{Das:2013oda}
		S.~Das,
		Phys. Rev. D \textbf{89}, no.8, 084068 (2014)
		\bibitem{Ashtekar:2008ay}
		A.~Ashtekar,
		J. Phys. Conf. Ser. \textbf{189}, 012003 (2009)
		\bibitem{Chakraborty:2019vki}
		S.~Chakraborty, D.~Kothawala and A.~Pesci,
		Phys. Lett. B \textbf{797}, 134877 (2019)
		\bibitem{Albareti:2012se}
		F.~D.~Albareti, J.~A.~R.~Cembranos and A.~de la Cruz-Dombriz,
		JCAP \textbf{12}, 020 (2012)
		\bibitem{Chakraborty:2023ork}
		M.~Chakraborty, A.~Bose and S.~Chakraborty,
		Phys. Scripta \textbf{98}, no.2, 025007 (2023)
		\bibitem{Choudhury:2021zij}
		S.~G.~Choudhury, A.~Dasgupta and N.~Banerjee,
		Int. J. Geom. Meth. Mod. Phys. \textbf{18}, no.08, 2150115 (2021)
		\bibitem{Albareti:2014dxa}
		F.~D.~Albareti, J.~A.~R.~Cembranos, A.~de la Cruz-Dombriz and A.~Dobado,
		JCAP \textbf{03}, 012 (2014)
		\bibitem{Blanchette:2021jcw}
		K.~Blanchette, S.~Das, S.~Hergott and S.~Rastgoo,
		doi:10.1142/9789811269776\_0356
		[arXiv:2110.05397 [gr-qc]].
		\bibitem{Casalderrey-Solana:2023zlg}
		J.~Casalderrey-Solana, D.~Mateos and A.~Serantes,
		[arXiv:2312.11643 [hep-th]].
		\bibitem{Belinsky:1970ew}
		V.~A.~Belinsky, I.~M.~Khalatnikov and E.~M.~Lifshitz,
		Adv. Phys. \textbf{19}, 525-573 (1970)
		\bibitem{Blanchette:2020kkk}
		K.~Blanchette, S.~Das, S.~Hergott and S.~Rastgoo,
		Phys. Rev. D \textbf{103}, no.8, 084038 (2021)
		\bibitem{Alsaleh:2017ozf}
		S.~Alsaleh, L.~Alasfar, M.~Faizal and A.~F.~Ali,
		Int. J. Mod. Phys. A \textbf{33}, no.10, 1850052 (2018)
		\bibitem{Crampin:2010}
		M.~ Crampin, T.~ Mestdag and W.~ Sarlet
		Z. Angew. Math. Mech. \textbf{90} (2010), 502-508
		
		\bibitem{Nigam:2016}
		K.~ Nigam, K.~ Banerjee
		``A Brief Review of Helmholtz Conditions" 
		arXiv:1602.01563
		\bibitem{Gotay:1983kvm}
		M.~J.~Gotay and J.~Demaret,
		Phys. Rev. D \textbf{28}, 2402-2413 (1983)
		\bibitem{Gotay:1980xk}
		M.~J.~Gotay and J.~A.~Isenberg,
		Phys. Rev. D \textbf{22}, 235-248 (1980)
		\bibitem{Lund:1973zz}
		F.~Lund,
		Phys. Rev. D \textbf{8}, 3253-3259 (1973)
		\bibitem{Jalalzadeh:2022dlj}
		S.~Jalalzadeh, A.~Mohammadi and D.~Demir,
		Phys. Dark Univ. \textbf{40}, 101227 (2023)
		[arXiv:2210.02629 [gr-qc]].
		\bibitem{Oppenheimer:1939ue}
		J.~R.~Oppenheimer and H.~Snyder,
		``On Continued gravitational contraction,''
		Phys. Rev. \textbf{56}, 455-459 (1939)
		\bibitem{Ziaie:2024saq}
		A.~H.~Ziaie, H.~Shabani and H.~Moradpour,
		Eur. Phys. J. Plus \textbf{139}, no.2, 148 (2024)
		\bibitem{Ratra:1987rm}
		B.~Ratra and P.~J.~E.~Peebles,
		Phys. Rev. D \textbf{37}, 3406 (1988)
		\bibitem{Chakraborty:2024nmg}
		S.~Chakraborty, S.~Mishra and S.~Chakraborty,
		Gen. Rel. Grav. \textbf{56}, no.7, 83 (2024)
		\bibitem{vs}
		Varun Sahni. Dark matter and dark energy. Lect. Notes Phys., 653:141–180, 2004.
		\bibitem{Ali:2014qla}
		A.~F.~Ali and S.~Das,
		Phys. Lett. B \textbf{741}, 276-279 (2015)
		\bibitem{Chakraborty:2023voy}
		M.~Chakraborty and S.~Chakraborty,
		Annals Phys. \textbf{457}, 169403 (2023)
			\bibitem{Chakraborty:2001za}
		S.~Chakraborty,
		Int. J. Mod. Phys. D \textbf{10}, 943-956 (2001)
		\bibitem{Chakraborty:2023yyz}
		M.~Chakraborty and S.~Chakraborty,
		Class. Quant. Grav. \textbf{40}, no.15, 155010 (2023)
		\bibitem{Chakraborty:2024dpx}
		M.~Chakraborty and S.~Chakraborty,
		Phys. Dark Univ. \textbf{46}, 101607 (2024)
	\end{thebibliography}
\end{document}